# Generalizing Thermal Transport in High-Contrast Metamaterials through Interfacial Fresnel Reflection

Seung Hyeon Ham[a], Yu Min Kim[a], In Hyeok Choi[b], and Jeong Woo Han [a,c*]

[a]*Department of Physics Education and Center for Quantum Technologies, Chonnam National University, Gwangju 61186, South Korea*
[b]*Department of Chemistry, Massachusetts Institute of Technology, Cambridge, Massachusetts 02138, USA*
[c]*Graduate School of Interdisciplinary Program for Photonic Engineering, Chonnam National University, Gwangju 61186, South Korea*

*Corresponding author: jwhan@chonnam.edu

**Abstract**

The rapid growth of generative AI has intensified the need for efficient heat dissipation in large-scale data centers. To control heat flow, thermal metamaterials with layered structures have been widely used, which impart the anisotropic properties of thermal conductivities. However, the conventional effective medium approximation (EMA) often fails to provide accurate predictions in systems with a high thermal conductivity contrast between adjacent layers embedded in a background medium. Here, we generalize the EMA by introducing two corrective coefficients that extend its validity to regimes where the conventional EMA was previously inapplicable, i.e., high-contrast thermal metamaterials with the background medium. Notably, one of these coefficients that we proposed has the same mathematical form as the Fresnel reflection coefficient in optics. This allows us to interpret the "reflection-like" behavior of heat flow as it penetrates adjacent layers with high thermal contrast. Our findings suggest that heat diffusion, traditionally viewed as a purely dissipative process, can be understood intuitively through the framework of ray optics.

## 1. Introduction

The rapid development of generative artificial intelligence (AI) models has driven the construction and operation of large-scale data-processing and storage centers, which consume enormous amounts of energy and consequently generate substantial heat [1-3]. Effective dissipation of this generated heat from computational hardware is therefore pivotal to efficient AI operation, such that controlling heat flow has emerged as an attractive academic and industrial research area more than ever [4-6]. Owing to high intrinsic thermal conductivity, metals are a proper substance for thermal management in computational devices. In addition, an anisotropic metallic system in terms of thermal conductivity is required in order to control the path of heat flow [7-11]; however, most metals in their natural state are primarily isotropic. Thus, artificially engineered composite materials in the form of a stacking structure with at least two metallic layers of different thermal conductivities, i.e., thermal metamaterials, have been widely used to impart anisotropic thermal properties [7-11]. For the consideration path of heat flow of thermal metamaterials, the effective medium approximation (EMA), which is based on the anisotropic thermal conductivity tensor, has been widely adopted [12-15]. The conceptual foundation of the EMA was inspired by transformation optics (TO), which utilizes coordinate transformations to determine the trajectories of electromagnetic waves in anisotropic media [8,16]. To extend these concepts to heat propagation, the governing Maxwell equations were substituted with Fourier's law combined with the continuity equation, leading to the heat flows from the high to low temperature with the diffusive motion [17,18]. However, because the original TO is rooted in Maxwell's equations, its mathematical validity is primarily confined to wave-like propagation. This leads to a fundamental limitation; while the mathematical forms appear similar, thermal transport is governed by a diffusive process rather than the wave propagation inherent in electromagnetic waves. Specifically, the EMA is no longer valid for the calculation of the heat flow, especially in the conditions in which (1) thermal conductivities ($\kappa_1$ and $\kappa_2$) of the adjacent layers are high contrast and (2) the thermal metamaterial is embedded in a background medium with thermal conductivity ($\beta$) comparable to $\kappa_1$ and $\kappa_2$, leading to heat leaking toward the background medium [19,20].

The intent of this study is to develop a generalized EMA that can be universally adopted for the realistic thermal metamaterials. We consider a thermal metamaterial composed of two metallic media with different thermal conductivities of $\kappa_1$ and $\kappa_2$. Our system is an *n*-layered structure which is embedded in the background material with the thermal conductivity of $\beta$. We systematically vary the ratio of $r=\kappa_1/\kappa_2$, $n$, and $\beta$ to quantitatively examine which conditions are suitable for the EMA. Similar to the prevailing views in other studies, our analysis shows that all three parameters are closely linked [7,14,21,22] . In particular, the heat flow in the systems with higher $r$, smaller $n$, and smaller $\beta$ tends to deviate from the prediction of the EMA. When heat flow crosses layers with high thermal conductivity contrast, reflection-like behavior analogous to optical reflection occurs. The conventional EMA does not account for this effect. We introduce a revised coefficient to incorporate it without modifying the EMA framework itself. Furthermore, while conventional EMA assumes that heat is directly induced into thermal metamaterials without the consideration of mediating materials, most of the realistic metamaterials are embedded within a background medium. To account for the background medium, we suggest the coefficient inspired by the principle of series connections in electronic circuits as well.

**2. Result**

Figure 1(a) shows the geometry of the thermal metamaterial used in this simulation study. It is composed of two metallic layers with distinct thermal conductivities, denoted by $\kappa_1$ and $\kappa_2$. The number of layers is indicated by $n$; here, we representatively plot seven layers ($n = 7$), as shown in Fig. 1(a). The layers are aligned diagonally at the angle of $\theta = 45^o$ with respect to the *x*-direction, which imparts anisotropic thermal conductivity to the thermal metamaterial. The background medium in which the layered structure is embedded has a thermal conductivity denoted by $\beta$. For easy interpretation, the layered structure and the background medium are both set to rectangular. The embedded thermal metamaterial is sandwiched between a heat source ($T_h$) and a heat sink ($T_c$), which have a bar-like structure. Both the heat source and the sink are maintained at constant high and low temperatures, thereby leading to the thermal equilibrium state with continuous heat flow. Without the layered structure, heat from the source would diffuse toward the sink along the surface normal. When heat enters the layered structure, its anisotropic properties, arising from the $\theta = 45^o$ alignment of the layered structure with respect to the background environment, cause the heat flow to bend. We define the effective bending angle of the heat flow as $\varphi$, which is determined by the line connecting the heat entry point to the corresponding exit point, as illustrated in Fig. 1(b) and calculated from Eq. (2). The bending of heat flow φ is accurately predicted by the EMA when the layered system can be assumed to be effectively isotropic, i.e., $\kappa_1 \approx \kappa_2$ (Fig. 1(b)). However, as the contrast between $\kappa_1$ and $\kappa_2$ increases, the deviation of the actual heat flow direction from the EMA prediction becomes more inaccurate (Fig. 1(c)) [7]. For the case of high contrast between the two layers (Fig. 1(d)), the heat flow is redistributed, leading to the reflection-like bounce back motion, when crossing the adjacent layers, which is in a similar manner to reflection in optics. These observations indicate that the applicability of the conventional EMA becomes limited as the conductivity contrast increases [7,23,24].

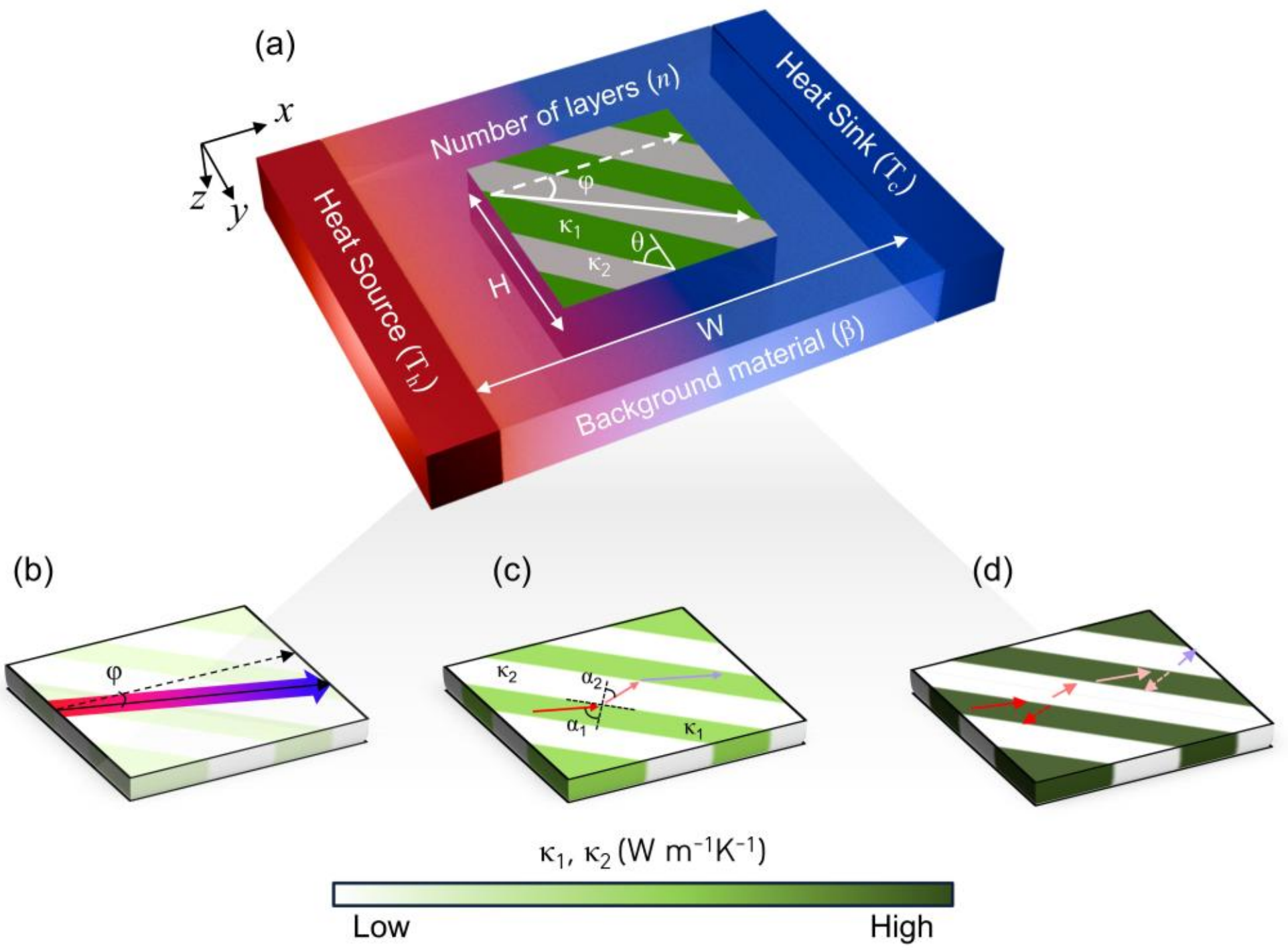

**Fig. 1.** (a) Schematic geometry of the thermal metamaterial used in the simulation study. The thermal conductivities of the two metallic layers and the background medium are denoted by $\kappa_1$, $\kappa_2$, and β, respectively, with *n* representing the number of layers. W is the width of the entire structure and H is the height of thermal metamaterial. The entire structure is sandwiched between a heat source ($T_h$) and a heat sink ($T_c$). The heat flow bending angle within the layered structure is denoted by φ. The metamaterial structure is aligned with $\theta = 45^o$. (b-d) Heat flow distributions under different conditions: (b) effectively isotropic, (c) partially anisotropic, and (d) strongly anisotropic. While cases (b) and (c) can be described by the conventional EMA method, case (d) exhibits reflection-like behavior at the interface due to the high contrast between $\kappa_1$ and $\kappa_2$, leading to the deviation from EMA in terms of the heat flow bending. $\alpha_1$ and $\alpha_2$ in (c) denote the incident and refractive angles of the heat flux.

To assess the validity of the conventional EMA under various conditions, we numerically compared the heat flow bending angle predicted by the EMA with that obtained from FEM as functions of *n*, *r*, and β for the thermal metamaterial shown in Fig. 1(a). These numerical simulations were performed by the finite element method (FEM) simulations (COMSOL Multiphysics). The layered structure has dimensions of $76 \times 76 \times 3$ mm$^3$, while the surrounding background medium measures 152×152×3 $mm^3$. The heat source and sink were maintained at 373 K and 273 K, respectively. To simulate a perfect heat absorber and eliminate the effects of heat radiation and convection from all boundaries in our simulation environments, which can potentially give rise to the artificial effect, we applied perfectly matched layers at all boundaries of interfaces for the background medium. To ensure the accuracy of the results independent of the mesh and the domain size, we demonstrated that the simulation results remain stable when the mesh and domain sizes are set below approximately 0.37 mm as the minimum element (see the Supplementary Information).

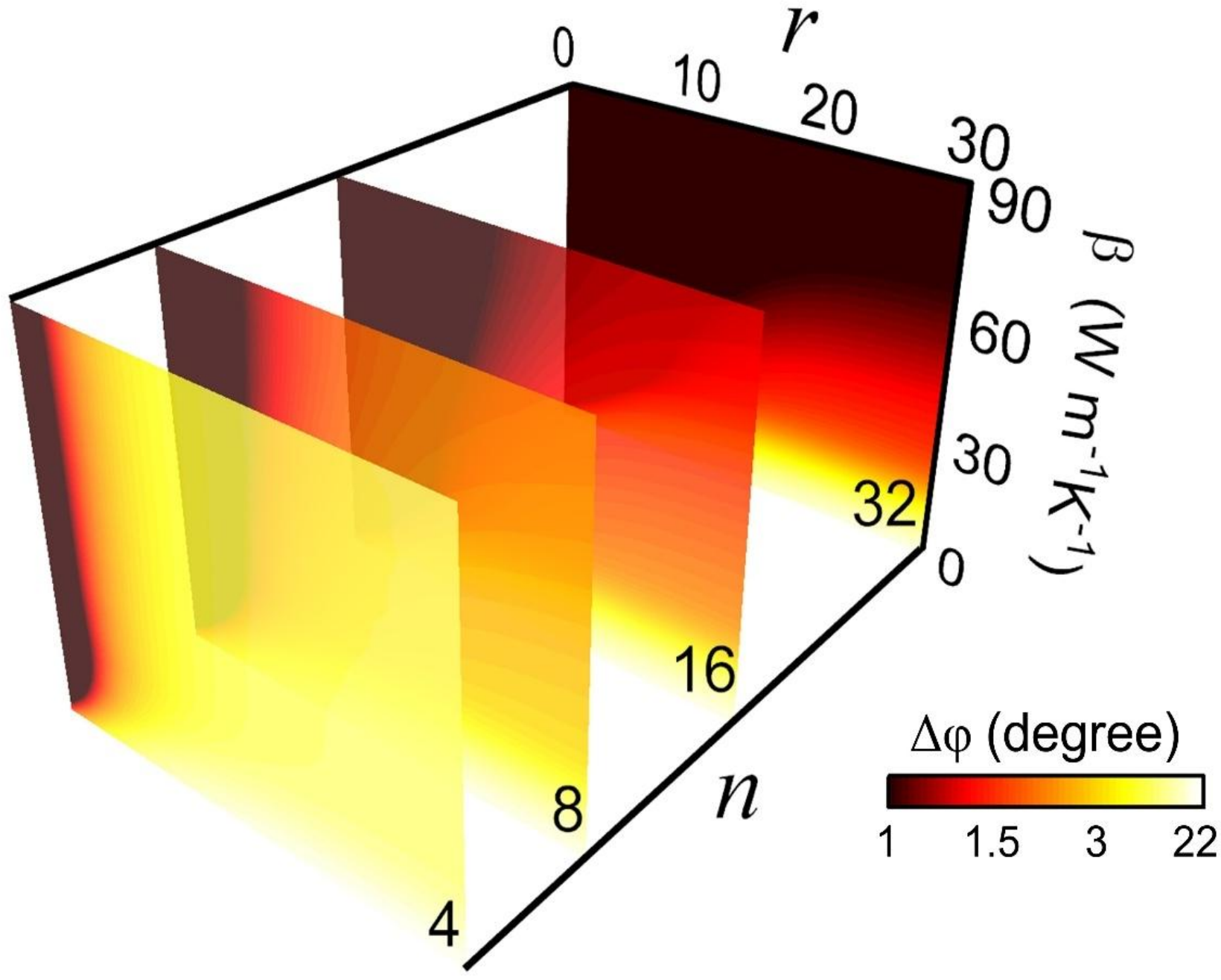

**Fig. 2.** Deviation of the heat-flow direction Δφ between the simulation results (FEM) and the prediction value from conventional EMA. *n* and *r* indicate the number of layers of the thermal metamaterial and the ratio of their thermal conductivities ($\kappa_1/\kappa_2$), respectively. β is the thermal conductivity of the background medium.

Figure 2 illustrates the deviation in the heat-flow bending angle (Δφ) on a logarithmic scale, defined as the difference between the analytical prediction of the conventional EMA and the numerical FEM results. For instance, $\Delta\varphi = 0^{o}$ means that the heat flow is perfectly described by the conventional EMA. The heat-flow bending angle (φ) predicted by the conventional EMA is given by [11,14,22]:

$$\varphi = \tan^{-1}\left(\frac{q_y}{q_x}\right) \tag{1}$$

where $q_x$ and $q_y$ are the *x*- and *y*- components of the heat-flux vector q, respectively, and q is computed by Fourier's law, i.e., $q = -\kappa\nabla T$. Since the thermal metamaterial considered in this paper is characterized by anisotropic thermal properties, κ is represented by the second-order tensor $\kappa_{ij}$, which is defined as follows:

$$\kappa_{ij}=\begin{pmatrix} \kappa_x \cos^2\theta + \kappa_y \sin^2\theta & (-\kappa_x+\kappa_y)\sin\theta\cos\theta \\ (-\kappa_x+\kappa_y)\sin\theta\cos\theta & \kappa_y\cos^2\theta + \kappa_x \sin^2\theta \end{pmatrix} \tag{2}$$

here θ denotes the angle between the interface and the *y*-axis (see Fig. 1(a)). Note that this κ tensor expression has been widely employed in the conventioanl EMA [7]. The parameters $\kappa_x$ and $\kappa_y$ are the effective thermal conductivities for the *x*- and *y*- directions, defined as $\kappa_x = (2\kappa_1\kappa_2 /(\kappa_1+ \kappa_2))$ and $\kappa_y = (\kappa_1+ \kappa_2)/2$, respectively. In the conventional EMA, $q_x$ and $q_y$ are calculated by Eq. (2) with the assumption that the temperature gradient ∇T is linearly distributed between the heat source, and hereafter φ is obtained by plugging them into Eq. (1). It is worth noting that the linear distribution of ∇T is not valid, especially when *r* is high contrast, leading to a localized non-uniform ∇T, resulting in eventually violation of the conventional EMA. The dependence of Δφ on *r*, *n*, and β is plotted to identify the parameter ranges over which the EMA can accurately predict φ (see Fig. 2). As can be seen, especially in the case of $n$ = 4, the EMA-valid region is initially restricted to small *r* values. However, as *n* increases, this valid regime extends toward larger *r*. One can interpret this trend as the fact that the temperature gradient redistribution becomes more pronounced when increasing *r*, which causes the deviation of the actual heat flow from that predicted by the conventional EMA. The influence of β becomes more significant at higher *n*. For instance, when $n$ = 32, the EMA accurately describes the heat flow even at $r$ = 30 ($\kappa_1$= 30 W $m^{-1}K^{-1}$ and $\kappa_2$=1 W $m^{-1}K^{-1}$), provided that β exceeds 60 W $m^{-1}K^{-1}$. Note that the bending angle of the heat flow φ is essentially determined by the ratio *r*, meaning that φ is the same regardless of the individual values of the thermal conductivities as long as *r* is the same. This observation suggests that the conventional EMA is not valid for systems in which heat leakage from the thermal metamaterial to the surrounding background medium is not negligible, which is consistent with the prevailing view.

To get the physical intuition about heat flow as it passes through the interface between the adjacent layers with high-contrast thermal conductivities ($\kappa_1$ and $\kappa_2$), we computed the temperature distribution in the thermal equilibrium state on a rectangular structure using Fourier's law. The simulation results are shown in Figs. 3(a)-(c), which correspond to $r$=100, 5, and 1, respectively. Note that $\kappa_2$ is set to 1 $Wm^{-1}K^{-1}$ and therefore *r* directly corresponds to $\kappa_1$. We assumed that the edges of the structure are maintained at 100 °C and 50 °C which serve as the heat source and the sink, respectively. The interface between the two layers is located at $x$ = 5, and the vertical axis represents temperature. For the case of $r$=100, the temperature is maintained at around 100 °C in the region with $\kappa_1$, but the temperature decreases linearly in the region of $\kappa_2$ (Fig. 3(a)). In contrast, the temperature falls linearly for the entire region with increasing thermal propagation distance when $r$=1 (Fig. 3(c)). The case of $r$=5 shows the intermediate behavior of the temperature distribution (Fig. 3(b)). This behavior can be understood from the fact that κ determines the speed of heat flow, causing the abrupt variation in the thermal distribution at the boundary with the change of κ. We provide the cross-section view at $y$=1,5, which illustrates the abrupt change in temperature distribution clearly (see Fig. 3(d)).

To quantitatively characterize the heat diffusion when crossing the layers with different thermal conductivities, i.e., $\kappa_1$ and $\kappa_2$, we integrated temperature ($A_1$: from $x$ = 0 to 5, and $A_2$: from $x$ = 5 to 10) for each layer along the *x* and *y* directions. Their ratio is defined by $A_2/A_1$. The thermal energy is determined by the product of the mass density, specific heat, and temperature. We herein set the mass density and the specific heat as constant values,

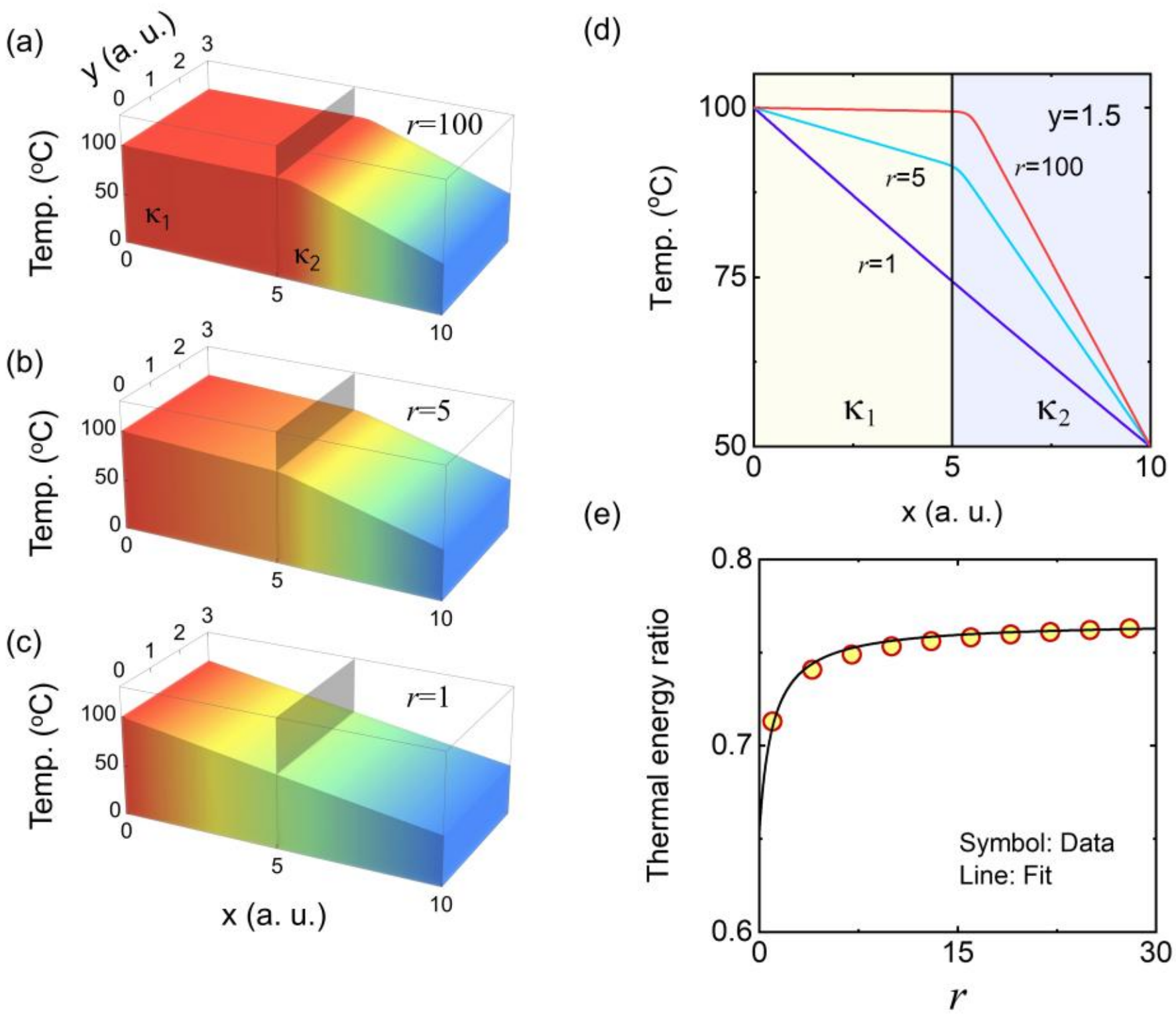

**Fig. 3.** Temperature distributions for the two-layer structure with thermal conductivity ratios of $r$=$\kappa_1/\kappa_2$=100 (a), 5 (b), and 1 (c). The interface between the two adjacent layers is located at $x$=5. (d) Corresponding temperature profiles along the $x$-direction at the cross section of $y$=1.5. (e) Thermal energy ratio, $A_2/A_1$, as a function of $r$. The symbols represent the integrated values of thermal energy (see the main text), while the solid line shows the empirical fit based on a mathematical form similar to the Fresnel reflection coefficient.

and hence this integrated temperature is directly proportional to the thermal energy stored in each of the materials. The thermal energy ratio ($A_2/A_1$) as a function of $r$ is presented in Fig. 3(e) as symbols. The obtained thermal energy ratio increases monotonically in the region where $r$ is less than 5 and slowly saturates thereafter. This trend is empirically well captured by the functional form $\alpha_1(r-1)/(r+1) + \alpha_2$, where $\alpha_1$ and $\alpha_2$ are the free fitting parameters, which turn out 0.56 and 0.71 for the best fit (see the line in Fig. 3(e)). The key observation is not the specific values of these free constants, but rather that the functional dependence on $r$ is well described by $(r-1)/(r+1)$. It is important to note that this functional format is identical to the Fresnel reflection coefficient [25,26]. The fact that the thermal energy ratio can be well described by the mathematical form of the Fresnel reflection coefficient indicates that the heat flow crossing the interface can be understood in a similar manner to reflection in optics. As a result, we employed the reflection coefficient $f(r)$, inspired by the reflection-type Fresnel coefficient, to treat the reflection-like behavior of the heat diffusion crossing the high-contrast two adjacent layers, as:

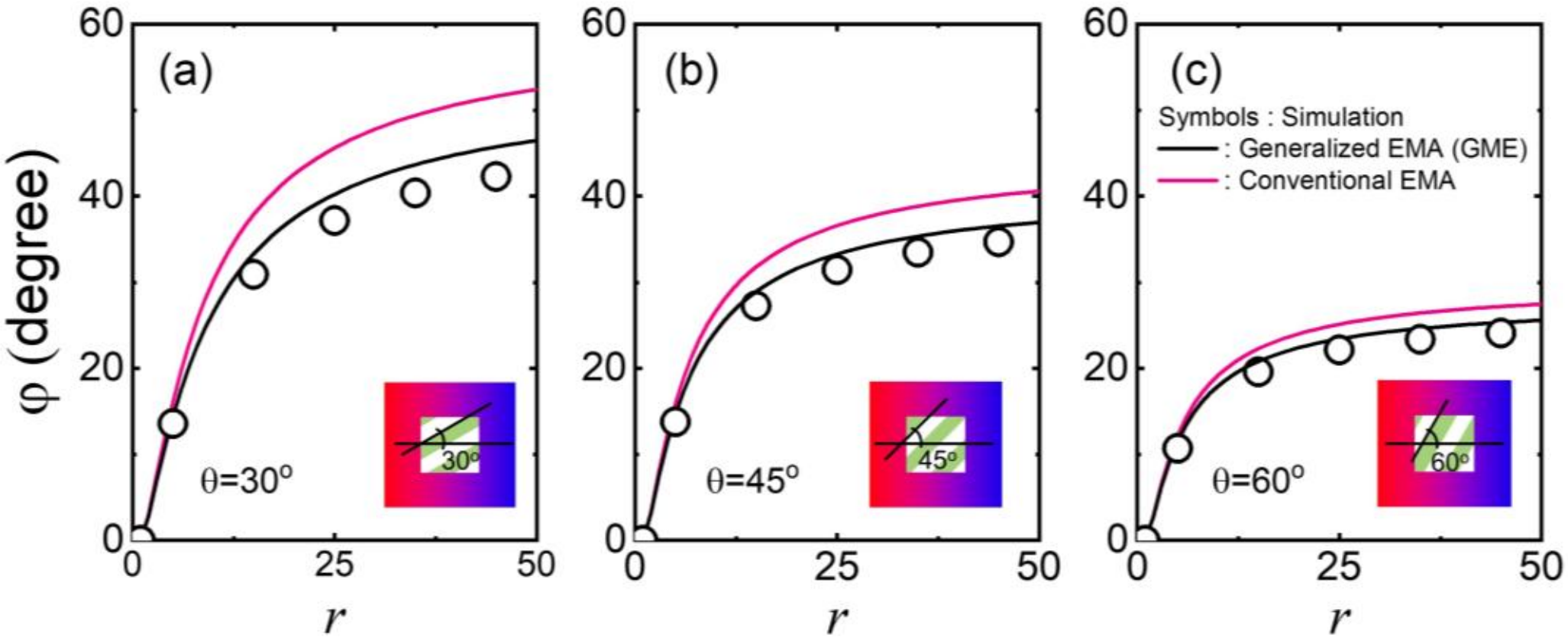

**Fig. 4.** Comparison of the heat-flow direction, φ, predicted by the generalized EMA (GMA) and the conventional EMA with numerical simulation results. The results for layer angles of θ=30 º (a), θ=45º (b), and θ=60º (c) are presented as a function of the thermal conductivity ratio *r* for a two-layer structure. Here, we set $n$=14, and β=30. Insets in Figs. 4(a)–(c) illustrate the geometry of the layered heat metamaterials.

$$f(r)=1-\frac{2}{n}\frac{r\sec\alpha_2-\sec\alpha_1}{r\sec\alpha_2+\sec\alpha_1} \tag{3}$$

where $\alpha_1$ and $\alpha_2$ are the incident and refraction angles for the heat flow, respectively (see Fig. 1(a)). They are expressed using the following trigonometric functions as $\sec\alpha_1=(1+((r+1)\cdot\tan\theta/2)^2)^{1/2}$ and $\sec\alpha_2=(1+((r+1)\cdot\tan\theta/2r)^2)^{1/2}$. It is important to note that the similarity in the mathematical form of $f(r)$ turns out to arise from the temperature gradient across the interface. The detailed mathematical derivation for $f(r)$ is provided in the Supplementary Information.

Most of the realistic cases, the thermal metamaterials are usually embedded within a background medium with thermal conductivity β, which mediates the heat exchange between the source/sink and the layered structure. This surrounding medium introduces an additional thermal resistance that the conventional EMA does not account for. To reproduce this effect thereby increasing the accuracy of EMA, we introduce a second corrective coefficient g(β), which is inspired by the principle of series connections in electronic circuits. The coefficient g(β) is written by:

$$g(\beta)=\frac{\beta}{\beta+\frac{2W}{m\pi H}K_{\mathrm{eff}}} \tag{4}$$

Here, W and H denote the distance between the heat source and heat sink and the vertical length of the metamaterial, respectively (see Fig. 1(a)). *m* is related to the spatial modulation profile of the temperature gradient in the background medium, induced by the embedded thermal metamaterial. This quantity arises during the derivation from Fourier's heat equation, and $m$=2 was chosen based on the actual temperature modulation for the geometry of the thermal metamaterial. $K_{\text{eff}}$ quantifies the degree of anisotropy in the metamaterial: it decreases as the anisotropy increases and increases as the anisotropy decreases. A detailed mathematical derivation and its physical interpretation are provided in the Supplementary Material. By multiplying $f(r)$ and $g(\beta)$ as the revised coefficients of the heat flux vectors ($q_x$ and $q_y$), one can generalize the EMA to be applicable to high-contrast thermal conductivity layers embedded in a background medium, as:

$$\varphi = \tan^{-1}\left( f(r) \cdot g(\beta) \cdot \frac{q_y}{q_x} \right) \tag{5}$$

Figures 4(a)-(c) show the angle ($\theta$)-dependent comparison of the heat-flow direction $\varphi$ between the conventional EMA and the generalized EMA, denoted by the red and black lines, respectively. Note that $\theta$ is defined as the rotation angle of the layered structure (see insets of Figs. 4(a)-(c) and Fig. 1(a)). We present three representative results for $\theta$ =30$^{\circ}$ (a), 45$^{\circ}$ (b), and 60$^{\circ}$ (c) as a function of the thermal conductivity ratio $r$. Here, we set $n$=14 and $\beta$=30. The main observations are as follows: (1) the generalized EMA predicts slightly smaller heat-flow direction angles $\varphi$ than the conventional EMA for all values of $r$, and even when $r$ is smaller; and (2) this discrepancy increases with $r$. The first trend mainly arises from the effect of the background material, represented by $g(\beta)$, since $f(r)$, representing the reflection-like behavior, plays a minor role when $r$ is smaller. This fact concomitantly suggests that the second trend can be attributed to the increasing importance of reflection-like behavior in heat diffusion as $r$ increases. The individual contributions of $f(r)$ and $g(\beta)$ to $\varphi$ are presented in the Supplementary Material. Taken together, these results indicate that reflection-like behavior, accounted for by the correction factor $f(r)$, is the primary source of the deviation from the conventional EMA at large $r$. In contrast, the overall reduction in $\varphi$ is primarily attributable to the background medium and is accounted for by $g(\beta)$. This agreement supports the physical validity of our approach and shows that the conventional EMA can be extended to systems comprising two high-contrast thermal-conductivity layers embedded in a background medium by incorporating the correction factors $f(r)$ and $g(\beta)$, respectively.

**3. Conclusion**

To control heat flow and dissipate thermal energy effectively, thermal metamaterials have been widely used to engineer anisotropic thermal conductivity. The effective medium approximation (EMA) has been developed to estimate the effective physical properties of thermal anisotropic materials. In this article, we generalize the EMA by introducing two coefficients. Conventional EMA may provide less accurate predictions for realistic thermal materials under two main conditions: (1) when the constituent layers have significantly different thermal conductivities, and (2) when the heat source and sink are not attached directly to the metamaterials. To address (1), which is particularly significant in high-contrast layered structures, we adopted a mathematical format similar to the Fresnel reflection coefficient in optics to account for "reflection-like" behavior in heat flow. Additionally, we introduced a coefficient inspired by series circuits to describe heat flow when embedded in a background medium. We confirm that our generalized EMA, incorporating these two coefficients, is in good agreement with the numerical simulations, even for metamaterials with both constraints (1) and (2). This suggests that our proposed generalized EMA extends the applicability of EMA to realistic thermal systems.

**CRediT authorship contribution statement**

**Seung Hyeon Ham:** Conceptualization, Data curation, Formal analysis. **Yu Min Kim:** Conceptualization, Data curation. **In Hyeok Choi**: Writing – review & editing. **Jeong Woo Han:** Visualization, Conceptualization, Project administration, Writing-Original draft, Writing-review & edition, Supervision.

**Acknowledgement**

This research was supported by the Regional Innovation System & Education (RISE) Glocal University 30 Program through the Gwangju RISE Center, funded by the Ministry of Education(MOE) and the Gwangju Metropolitan City, Republic of Korea.((2026-RISE(Glocal University 30)-05-011).

**Declaration of competing interest**

The authors declare that they have no known competing financial interests or personal relationships that could have appeared to influence the work reported in this paper.

**Data availability**

Data will be made available on request